\documentclass[natbib,graybox]{svmult}

\usepackage{mathptmx}       
\usepackage{helvet}         
\usepackage{courier}        
\usepackage{type1cm}        
\usepackage{makeidx}         
\usepackage{graphicx}        
\usepackage{multicol}        
\usepackage[bottom]{footmisc}

\usepackage{aas_macros}

\makeindex             

\def\mst{\mbox{$M_{\star}$}}

\def\Re{\mbox{$R_{\rm e}$}}
\def\Msun{\mbox{$M_\odot$}}

\def\ML{\mbox{$M/L$}}

\def\Mvir{\mbox{$M_{\rm vir}$}}
\def\cvir{\mbox{$c_{\rm vir}$}}
\def\atlas3d{ATLAS$^{\rm 3D}$}
\def\lsim{\mathrel{\rlap{\lower3.5pt\hbox{\hskip0.5pt$\sim$}}
    \raise0.5pt\hbox{$<$}}}                
\def\gsim{~\rlap{$>$}{\lower 1.0ex\hbox{$\sim$}}}

\begin{document}

\title*{Systematic variation of central mass density slope in early-type galaxies}
\author{C.~Tortora,  F.~La~Barbera,  N.~R.~Napolitano, A.~J.~Romanowsky,
I.~Ferreras, \and R.~R.~de Carvalho} \institute{C.~Tortora
\email{ctortora@na.astro.it}, F.~La~Barbera, N.R.~Napolitano \at
INAF -- Osservatorio Astronomico di Capodimonte, Salita
Moiariello, 16, 80131 - Napoli, Italy, \and A.J.~Romanowsky \at
Department of Physics and Astronomy, San Jos\'e State University,
San Jose, CA 95192, USA, \at University of California
Observatories, 1156 High Street, Santa Cruz, CA 95064, USA, \and
I.~Ferreras \at Mullard Space Science Laboratory, University
College London, Holmbury St Mary, Dorking, Surrey RH5 6NT, \and
R.~R.~de Carvalho \at Instituto Nacional de Pesquisas Espaciais /
MCTI Av. dos Astronautas 1758, Jd. Granja S$\rm \tilde{ã}$o Jos\'e
dos Campos - 12227-010 SP} \maketitle

\abstract{We study the total density distribution in the central
regions ($\lsim \, 1$ effective radius, \Re) of early-type
galaxies (ETGs), using data from the SPIDER survey
(\citealt{SPIDER-I}). We model each galaxy with two components
(dark matter halo + stars), exploring different assumptions for
the dark matter (DM) halo profile, and leaving stellar
mass-to-light (\mst$/L$) ratios as free fitting parameters to the
data. For a \cite{NFW96} profile, the slope of the total mass
profile is non-universal. For the most massive and largest ETGs,
the profile is isothermal in the central regions ($\sim \Re/2$),
while for the low-mass and smallest systems, the profile is
steeper than isothermal, with slopes similar to those for a
constant-\ML\ profile. For a concentration-mass relation steeper
than that expected from simulations, the correlation of density
slope with mass tends to flatten. Our results clearly point to a
``non-homology'' in the total mass distribution of ETGs, which
simulations of galaxy formation suggest may be related to a
varying role of dissipation with galaxy mass.}

\vspace{1cm}

\begin{figure*}[!]
\centering
\includegraphics[scale=.54]{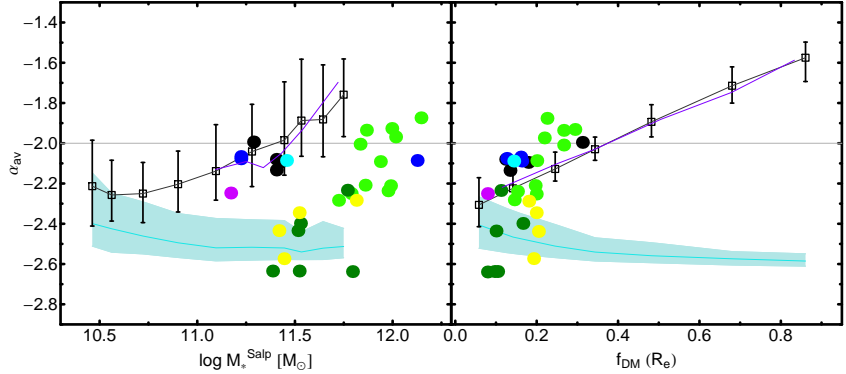}
\caption{Comparison of total mass density slope trends with
predictions for simulated galaxies from \citet{Remus+13}. Open
squares and error bars are median and 16--84th percentile trends
for our reference NFW halo model and a variable IMF,
\citet{Remus+13} adopt a \citet{Salpeter55} IMF. Left and right
panels plot the average slope (according to the definition in
\citealt{Remus+13}) vs. Salpeter-based stellar mass (derived from
SPS modelling) and vs. DM fraction within 1 \Re\ (derived from the
variable IMF models), respectively. Purple lines are the observed
trends for the subsample of SPIDER ETGs with Salpeter-like IMF
normalization. The cyan line and shaded regions mark median and
16--84th percentile slopes for the stellar mass distribution only.
Simulations consist of binary (spiral-spiral and
spiral-elliptical) mergers of different mass ratios and remnants
of multiple mergings from cosmological simulations. Dots with
different colours are simulated galaxies from \citet{Remus+13}:
black, blue, cyan and pink dots correspond to idealized single
binary mergers, while light-green, dark-green, and yellow dots are
for mergings systems drawn from cosmological simulations. See
\citet{Tortora+14_DMslope} for further details about colour
coding.} \label{fig:1}
\end{figure*}

We use a sample of $\sim 4300$ local ETGs from the SPIDER survey
(\citealt{SPIDER-I}) in the redshift range 0.05-0.1.  Our analysis
extends the range of galaxy stellar mass ($\mst$) probed by
gravitational lensing, down to $\sim 10^{10}\, \rm \Msun$. We
model the observed velocity dispersion of each individual galaxy
using spherical isotropic Jeans equations. The stellar density is
provided by a S\'ersic fit of the photometric data from SDSS-DR6
and UKIDSS-LAS-DR4 (see \citealt{SPIDER-I} for further details).
DM mass is modelled with a suite of profiles. Further details are
provided in \cite{Tortora+14_DMslope}.


Independently of the DM halo adopted, we find that the Initial
Mass Function (IMF) becomes bottom-heavier than a ``standard''
\cite{Chabrier01} distribution in high- relative to low-mass ETGs
(\citealt{LaBarbera+13_SPIDERVIII_IMF,
TRN13_SPIDER_IMF,Tortora+14_DMslope}). One of the most important
results of our work is that total mass density slope in ETGs is
not universal, assuming a NFW halo model (\citealt{NFW96}) with
standard recipes for the virial to stellar mass relation
(\citealt{Moster+10}) and \cvir--\Mvir\ relation
(\citealt{Maccio+08}). Our results suggest that total mass density
slope gets shallower with galaxy mass, galaxy size and central DM
fraction (see Fig.~\ref{fig:1}). In more detail, we find that
low-mass (small) ETGs have slopes consistent with those for
constant-\ML\ profiles (see cyan shaded region in
Fig.~\ref{fig:1}), while massive (large \Re) systems have a nearly
isothermal density slope ($=-2$) or shallower, consistent with
gravitational lensing results (\citealt{Auger+10_SLACSX}). The
trends of mass density slope are consistent with independent
results from the literature
(\citealt{Dutton_Treu14,Humphrey_Buote10}). The trends of mass
density slope are the same for NFW and contracted-NFW models, and
do not change when assuming a fixed virial mass (and
concentration) for all galaxies. When adopting a \cite{Burkert95}
profile, the slope tends to be more constant as a function of all
galaxy parameters explored. However, for the most massive ETGs,
the ``light'' haloes described by Burkert models seem to be
rejected by lensing results (\citealt{Auger+10_SLACSX}). Using
\cvir\ -- \Mvir\ relation from observations (\citealt{Leier+12})
rather than simulations affects significantly some trends of
density slope with galaxy parameters. In particular, while the
slope keeps increasing with galaxy radius  also for such
``high-concentration'' models, the trends with mass become flatter
in this case.

Our results draw a picture whereby {\it the total mass density
profile in the central regions of ETGs is ``non-homologous''},
approaching a constant-\ML\ distribution at low mass -- where
stars dominate the total mass budget in the ETGs center --, and an
isothermal profile in the most massive ETGs, whose central regions
are more DM dominated. To understand the implications of our
findings in the framework of galaxy assembly, in Fig.~\ref{fig:1}
we have also compared our results  with predictions from N-body
simulations (\citealt{Remus+13}). We find that both observations
and simulations predict an increase of the total mass density
slope with galaxy mass. This comparison indicates that black hole
growth and feedback are fundamental ingredients during the
formation of ETGs, as only simulations including these processes
are able to reproduce the mass density slopes, DM fractions, and
stellar masses in the central regions of ETGs. We argue that this
trend is because gas dissipation has been more important during
the formation of low-, relative to high-, mass ETGs. Thus, a steep
profile is due to the formation of new stars inwards, as the gas,
dissipating its kinetic energy, falls into the galaxy central
regions, while gas-poor mergers tend to make the slopes
isothermal.

\begin{acknowledgement}
CT has received funding from the European Union Seventh Framework
Programme (FP7/2007-2013) under grant agreement n. 267251
``Astronomy Fellowships in Italy'' (AstroFIt)
\end{acknowledgement}

%
%
%


\end{document}